\documentclass[12pt,preprint]{aastex}
\begin{document}
\title{The Highly Relativistic Kiloparsec-Scale Jet of the Gamma-Ray Quasar 0827+243}
\author{Svetlana G. Jorstad\altaffilmark{1,2} and Alan P. Marscher\altaffilmark{1}}
\altaffiltext{1}{Institute for Astrophysical Research, Boston University,
    725 Commonwealth Ave., Boston MA 02215}
\altaffiltext{2}{Sobolev Astronomical Institute, St. Petersburg State University,
    28 Universitetskij pr., St. Petersburg 198504 Russia}
\begin{abstract}
We present {\it Chandra} X-ray (0.2-8 keV) and {\it Very Large Array} radio (15 and 5 GHz)
images of the $\gamma$-ray bright, superluminal quasar 0827+243. The X-ray jet
bends sharply---by $\sim 90^\circ$, presumably amplified by projection effects---5$''$
from the core. Only extremely weak radio emission is detected between the nuclear
region and the bend. The X-ray continuum spectrum of the combined emission of the knots is
rather flat, with a slope of $-0.4\pm 0.2$, while the 5-15 GHz spectra are steeper for
knots detected in the radio. These characteristics, as well as non-detection of the
jet in the optical band by the {\it Hubble Space Telescope}, pose challenges to
models for the spectral energy distributions (SEDs) of the jet features.
The SEDs could arise from pure synchrotron emission from either a single or dual population of
relativistic electrons only if the minimum electron energy per unit mass
$\gamma_{\rm min} \gtrsim 1000$. In the case of a single population, the
radiative energy losses of the X-ray emitting electrons must be suppressed owing to
inverse Compton scattering in the Klein-Nishina regime, as proposed by Dermer \& Atoyan.
Alternatively, the X-ray emission could result from  inverse Compton scattering of the
Cosmic Microwave Background photons by electrons with Lorentz factors as low as
$\gamma \sim 15$. In all models, the bulk Lorentz factor of the jet flow
$\Gamma\gtrsim 20$ found on parsec scales must continue without
substantial deceleration out to 800~kpc (deprojected) from the nucleus, and the magnetic
field is very low, $\lesssim 2$ $\mu$G, until the bend. Deceleration does appear to occur
at and beyond the sharp bend, such that the flow could be only
mildly relativistic at the end of the jet.  Significant intensification of the magnetic
field occurs downstream of the bend, where there is an offset between the projected
positions of the X-ray and radio features.
\end{abstract}
\keywords{galaxies: quasars: general --- galaxies: jets ---
galaxies: quasars: individual (\objectname{0827+243}) --- X-rays: galaxies --- radio
continuum: galaxies}

\section{Introduction}

Although the existence of X-ray emission in the jets of active galactic nuclei
was known prior to the launch of the {\it Chandra} X-ray observatory, it was
surprising to discover that a substantial fraction of observed kiloparsec-scale
jets have detectable X-ray counterparts \citep[e.g.,][]{SAM04,Schw03,GJ03}. In
some cases---notably the Fanaroff-Riley I radio galaxies
\citep[e.g.,][]{W01} and some quasars \citep{SAM04}---the multiwaveband spectra indicate
that the X-ray emission is synchrotron radiation by electrons with energies
$\gtrsim 1$ TeV. In others, inverse Compton scattering of the Cosmic Microwave Background
(CMB) off $\sim 0.1$ GeV electrons in the jet (IC/CMB) provides an alternative
X-ray emission mechanism \citep{Tav00,Cel01,SAM04}. The
synchrotron self-Compton process may be important in some
condensed regions with high densities of relativistic electrons
\citep[see, e.g.,][]{HK02}.

IC/CMB emission should be most pronounced in high-redshift objects with
jets whose flow velocities remain highly relativistic out to scales of tens
or hundreds of kiloparsecs, as long as one of the two opposing jets
points almost directly along the line of sight. In this context
$\gamma$-ray blazars, whose jets display faster apparent speeds
than the general population of compact flat spectrum radio sources
\citep{J01,KL04}, are prime candidates for strong IC/CMB emission. Detection of
IC/CMB X-rays provides a determination of the Doppler factor on kiloparsec
scales \citep[see Appendix A and][]{DA04}, which can then be compared with that
measured on parsec scales.
A number of kiloparsec scale jets show a decrease in the ratio
of X-ray to radio intensity with distance from the core \citep{MH01,
Siem02,SAM04}. This could result from gradual decrease of the bulk Lorentz factor
on kiloparsec scales \citep{GK04}. Such a model can be tested only
in the case of observation of multi-knot jet structure
at two  or more frequencies, which are rarely available despite the high
sensitivity of Chandra observations.    
For these reasons, the detected X-ray/radio jet of the $\gamma$-ray bright quasar
0827+243 \citep[$z = 0.939$; ][]{HB93}, which has apparent superluminal motion
on millarcsecond scales as fast as $\beta_{\rm app} \approx 22c$ \citep{J01}, represents
an excellent laboratory for studying the physical conditions in kiloparsec-scale jets.
\footnote[4]{An incorrect redshift was used by \citet{J01}, who
followed the value given in \citet{H99}. The apparent velocity given
uses the correct redshift of 0.939. Proper motions are converted to speeds under
a standard Friedmann cosmology with $H_0 = 70$ km s$^{-1}$ Mpc$^{-1}$ and
$q_0=0.1$. The conversion of angular to linear scale projected on the sky
is then $1'' = 7.2$ kpc.}

\section{Observations and Data Reduction}

We have observed the quasar 0827+243 with the {\it Chandra} Advanced CCD Imaging Spectrometer
(ACIS-S) at 0.2-8 keV and with the {\it Very Large Array} (VLA) interferometer of the
National Radio Astronomy Observatory (NRAO)
at 15.0~GHz (B array) and at 4.9~GHz (A array). This observational configuration
provides similar resolution ($\sim 0''\kern -0.35em .5$) at all three
frequencies. We have also retrieved an optical image of the quasar from the
{\it Hubble Space Telescope} archive.

\subsection{X-Ray Observations}
We observed the quasar for 18.26 ks on 2002 May 10 with the ACIS-S detector of the
{\it Chandra} X-Ray Observatory, using the back-illuminated S3 chip.
In order to minimize the effect of pileup of the emission from the core,
we used a 1/8 subarray to reduce the frame time to 0.4~s.
The X-ray position
of the quasar ($08^h30^m52^s\kern -0.35em .089$,$+24^\circ10'59''\kern -0.35em .67$, J2000)
agrees very well with the radio position
($08^h30^m52^s\kern -0.35em .086$,$+24^\circ 10'59''\kern -0.35em .82$).
We used version 3.0.2 of the CIAO software and the version 2.26 of the CALDB calibration
database for the data analysis.
We generated a new level 2 event file to apply an updated gain map, to randomize the 
pulse height amplitude (PHA) values, and to remove the pixel randomization.
We extracted the background by using four circular regions with radius $2''\kern -0.35em .5$
each on the east, north, west, and south sides at a distance $\sim 10''$ from the core,
avoiding the readout streaks. We applied the pileup model in Sherpa to determine
the pileup fraction of the core, 5.5\%.

Figure 1 shows the smoothed image of the quasar in the range 0.2-8~keV. The image has a
bright core and a jet that extends out to $\sim 6''\kern -0.35em .5$. At about $5''$ the
jet, as defined by the emission, executes a sharp bend by $\sim 90^\circ$.
The first section of the jet, between the core and the bend, is aligned with the direction
of the parsec scale jet \citep{J01}. The X-ray structure can be represented by three circular
components ($C1, C2$, and $C3$) and an elliptical component ($C4$)
with axial ratio $\sim 0.6$. Table 1 gives extracted counts for the core, knots, and
background. All knots are detected at a level of 6$\sigma$ or higher
relative to the background estimate.
Table 1 shows that the hardness ratio peaks in the brightest jet component.
We calculated the redistribution matrix and auxiliary response files for the core and jet regions
using the thread {\it Extract ACIS Spectra for Pointlike Source and Make
RMFs and ARFs}. We fit the spectral data for the core and jet in the range 0.2-8~keV
in Sherpa by a power-law model with fixed
Galactic absorption $N_H=3.62\times 10^{20}$ corresponding to
the direction of the quasar, obtained with routine COLDEN based on the data of
\citet{DL90}. The best-fit photon indices are very similar
for the core and extended jet, 1.34$\pm$0.02 and 1.4$\pm$0.2, respectively.
This is consistent with the flat X-ray spectra of bright jet knots found previously in
other quasars \citep{SAM04,Siem02}. The jet emits a fraction $\sim$0.01 of the apparent core
luminosity $L_{core}=(1.16\pm 0.07)\times 10^{46}$~erg~s$^{-1}$ over the energy
range 0.2-8~keV.
We calculated the flux of the individual jet components at 1~keV (see Table 2) after fixing the
jet power-law photon index at 1.4 and the Galactic absorption as indicated above.
 
We used the {\it Chandra} Ray Tracer facility to produce a point spread function (PSF) to
match the core region. The PSF was calculated for the position of the core,
its spectrum at 0.2-8~keV, and the exposure time. The PSF was projected onto the detector
using a thread in the MARX program. A subpixelated image of the PSF was created by
following the thread {\it Creating an Image of the PSF} in CIAO. The resulting
image is displayed in Figure 1.

\subsection{Radio Observations}

We performed observations with the {\it Very Large Array} (VLA) of the National Radio Astronomy
Observatory (NRAO) at 15 GHz (central frequency = 14.9649 GHz,
bandwidth = 50 MHz) in B array on 2002 August 19, and at 5 GHz
(central frequency = 4.86010 GHz, bandwidth = 50 MHz) in A array
on 2003 August 28. We edited and calibrated the data using
the Astronomical Image Processing System (AIPS) software provided by NRAO.
The resulting images are shown in Figure 2. Although the resolution of the VLA
observations $\sim 0''\kern -0.35em .4$, we convolved the images with a circular Gaussian
beam of FWHM = $0''\kern -0.35em .5$ in order to match the resolution of the
{\it Chandra} images. We modeled the total and polarized intensity images with circular
Gaussian components using the software Difmap \citep{Difmap} and IDL (by Research
Systems, Inc.), respectively.

The core has an inverted radio spectrum and is linearly polarized at 5~GHz at a level of
$4.5\pm0.5$\% with electric vector position angle (EVPA) of $43^\circ$, roughly
transverse to the main parsec-scale jet direction \citep{J01}.
At 15~GHz the core has a more modest fractional polarization (1.8$\pm$0.7\%)
with EVPA = $116^\circ$. The latter is close to the direction and EVPA of the innermost jet
just outside the parsec-scale core before it bends toward the kiloparsec-scale
structure \citep{APM02,J01}. The difference in polarization at
the two frequencies might be due to variability
of the core radio emission, since the VLA observations were separated by 1~yr.

The images reveal very interesting structure of the
kiloparsec-scale jet: (i) knot $B1$, $\sim 0''\kern -0.35em .4$ from the core,
has a position angle corresponding to that of the main parsec-scale jet \citep{J01};
(ii) despite high dynamic range (RMS $<$ 0.1~mJy), neither the 5~GHz
nor 15~GHz image shows significant radio emission between $\sim 0''\kern -0.35em .5$ and 5$''$
from the core where the brightest X-ray features are observed; and (iii) the large-scale radio jet
starts beyond 5$''$ and is roughly perpendicular to the parsec-scale jet as projected on the
sky, corresponding to the direction of the second section of the extended X-ray structure.
The 5-15 GHz spectral index $\alpha$ (flux density $S_\nu\propto\nu^{-\alpha}$) of the knots
generally increases with distance from the core, from $\sim 0.7$ ($B1$) to $\sim 1.1$ ($C5$).
However, the brightest radio feature, $C6$, at the edge of the kiloparsec-scale
jet has a radio spectral index $\alpha \sim 0.8$ and is highly
polarized ($20\pm 5$\%), with EVPA parallel (magnetic field perpendicular) to the local jet
direction. Curiously, feature $C3$ near the bend is detected at 5 GHz but not at 15 GHz, which
requires an extremely steep spectrum, $\alpha \geq 1.5$.

\subsection{Optical Data}

A field containing 0827+243 was observed by \citet{OPT02} with the Wide Field Planetary
Camera 2 (WFCP2) of the {\it Hubble Space Telescope} (HST) as a part of a
survey of QSO-absorbing galaxies at intermediate redshifts. We retrieved
the images of this field from the HST archive. The images
were obtained in the F702W ($\lambda_{\rm eff}=708.14$ nm) filter on 1995 May 29 with
a total exposure of 4.6~ks, and calibrated by the HST standard processing pipeline.
We used the European Southern Observatory image reduction software MIDAS to analyze the
images (draw profiles and measure magnitudes). We applied the absolute flux
calibration given in the headers of the images and in the WFPC2
Instrument Handbook
to convert the instrumental magnitudes to fluxes.

Figure 3
traces the optical profiles along the X-ray jet ($a$) and two rays, $b$ and $c$,
in the counterjet direction and perpendicular to the jet-counterjet line, respectively.
The {\it a} profile shows a slight increase in the number of counts in the region of
the brightest X-ray component $C2$, but at an insignificant level.
The estimated $1\sigma$ upper limits are $2.5\times 10^{-30}$ erg~cm$^{-2}$~s$^{-1}$~Hz$^{-1}$
for $C2$ and $2.1\times 10^{-30}$ erg~cm$^{-2}$~s$^{-1}$~Hz$^{-1}$ for
$C1, C3, C4, C5,$ and $C6$.

\section{Properties of the X-Ray/Radio Jet Emission}

The most striking features of the combined X-ray and radio images (Fig. 4) are the strong
bend in the kiloparsec-scale jet and the absence of detected radio emission over much of
the X-ray bright section.
The X-ray emission in 0827+243 extends about $5''\kern -0.35em .5$ from the core,
then executes a bend of $\sim 90^\circ$ in projection on the sky. Beyond this point, the
X-ray jet continues for about $2''$. Outside the nuclear region, the radio emission first
rises above the noise level only near the bend. The direction of the inner X-ray jet is within
$5^\circ$ of that of the parsec-scale radio jet, where \citet{J01} have measured
apparent speeds as high as $\beta_{\rm app}=22\pm 2c$. This apparent speed implies
that the bulk Lorentz factor
$\Gamma \geq 22$, with an optimal viewing angle $\theta \leq 2^\circ\kern -0.35em .5$.
Using the least extreme values of $\Gamma$ and $\theta$ provides an estimate of the Doppler
factor on parsec scales, $\delta \equiv [\Gamma(1-\beta\cos\theta)]^{-1} \sim \Gamma \approx 22$,
where $\beta$ is the actual bulk velocity in units of $c$.

Figure 5 displays the inner region of the X-ray/radio jet after subtracting the PSF
from the core. Two new components close to the core, $B1$ and $B2$, are apparent
in the X-ray jet (of these only knot $B1$ is detected in the radio), aligned to
within $2^\circ$ of the main parsec-scale structure. When projection effects are considered,
this indicates an essentially linear continuation from
parsec to kiloparsec scales up to the sharp-angle bend. Table 2 lists the parameters of the
kiloparsec-scale jet features at 1~keV and 5~GHz. Figure 6 shows the X-ray and radio
intensity profiles along the jet axis as indicated in Figure 1 after core subtraction,
as well as the spectral index variation along the jet.

The profiles displayed in Figure 6 show clearly the fading of the X-ray emission
with distance from the nucleus while the radio
intensity increases. Similar X-ray/radio intensity-ratio gradients are found in several
other quasars (3C~273,
\citealt{MH01}; PKS~1127$-$145, \citealt{Siem02}; PKS~1136$-$135 and
PKS~1510$-$089, \citealt{SAM04}), while in the majority of jets detected by {\it Chandra}
the X-ray knots coincide with bright radio features \citep{SAM04}. Note that
3C~273, PKS~1127$-$145, PKS~1510$-$089, and 0827$+$243 are all $\gamma$-ray bright
quasars with high apparent speeds ($\gtrsim 10c$) on parsec scales \citep{J01}.

Because of limited photon statistics, the angular sizes determined from the X-ray image
are uncertain by $\sim 50$\%. Similarly, the offsets of the X-ray and radio
intensity apparent in Figs. 4 and 5 in features $C3$ and $C4$ may not be real.
The presence of any X-ray photons outside the core region {\it is} highly significant,
but the absence of such photons near the periphery of a knot could be due to either a
genuine lack of emission or insufficient exposure time.

In order to produce an apparent bend of $90^\circ$, the jet would need to change its
intrinsic direction by an angle at least equal to the
initial viewing angle. Since this is only $\sim 2^\circ\kern -0.35em .5$, the intrinsic
change in trajectory can be quite modest. The bend could be caused by deflection of the
jet flow by a cloud, in which case one would expect a standing shock wave to
compress and heat the jet plasma while it decelerates the bulk motion. The observed
drop in intensity around the bend (see Fig. 6) implies that the decrease in Doppler beaming
more strongly influences the observed emission than does the compression and heating.

An alternative scenario for the apparent bending is that the jet flow has always been
straight, but the direction of the ``nozzle'' changed. That is, about $4\times 10^6$~yr
in the past the jet might have pointed from the
nucleus to knot $C6$, then steadily swung until it reached its current direction about
$2.5\times 10^6$ yr ago. In this case, knots $C4$, $C5$, and $C6$ could no longer
be energized by the jet. The radiative lifetime of electrons emitting at a frequency
$\nu_{\rm s,GHz}$ GHz is
\begin{equation}
t_{\rm loss} \approx 1.3\times 10^6 ({{B_{\mu G}\delta}\over{\nu_{\rm s,GHz}(1+z)}})^{1/2}
[B_{\rm \mu G}^2 + 11(1+z)^4\Gamma^2]^{-1}~~{\rm yr},
\end{equation}
where the quantities on the right-hand side are all defined in Appendix A. This includes
energy losses to both synchrotron radiation and IC/CMB scattering in the Thomson
limit. For the magnetic fields we derive in the next section,
the lifetimes of electrons (times the bulk Lorentz factor $\Gamma$ to translate to
the rest frame of the host galaxy) radiating at 15 GHz are
considerably shorter than the estimated age of knot {\it C6}, $\gtrsim 4\times 10^6$ yr.

\section{The Nature of X-ray Emission in the Large-Scale Jet}

The most popular models to explain the observed spectral energy distribution (SED) of
features on kiloparsec scales involve (1) synchrotron radiation from a single population
\citep{MH01,HK02,DA02} or dual population \citep{HK02,AD04,SAM04} of relativistic
electrons, and/or (2) inverse-Compton scattering off the Cosmic Microwave Background
\citep[IC/CMB,][]{Tav00, Cel01}. The detection of several X-ray knots at different distances
from the core allows us to constrain the possible emission processes, and then
to derive key physical parameters and how they vary as a function of distance from
the nucleus.

Measurements of the radio synchrotron emission determine the value of the product $B\delta$, 
or an upper limit for knots that are not detected at either 5 or 15 GHz, through eq. (A3). 
This product depends on the unknown quantities $k$, $A$, and $\phi$, which represent the ratio 
of proton to electron energy density, the ratio of particle to magnetic energy density, 
and volume filling factor, respectively. However, these are raised to a small power, 
hence only unlikely, extreme values of these parameters could affect greatly
the derived value of $B\delta$.
The models for the X-ray emission from the jet of 0827+243 must therefore be consistent with 
the values or limits to $B\delta$ dictated by the observations.

Figure 7 shows the observed spectral energy distributions (SEDs) for the three
most prominent regions in the jet: knot $C2$, with the brightest
X-ray emission and non-detection at optical and radio frequencies,
$C4$, with measured X-ray and radio fluxes and optical upper limit,
and $C6$, detected only in the radio. The X-ray spectral index,
$\alpha_{\rm x} = 0.4\pm 0.2$, is significantly flatter than the radio spectral indices, 
$\alpha_{\rm r} =$ 0.7-1.1.
(As discussed above, there are only enough photon counts to determine the X-ray spectral index
for all the knots combined. Therefore, it is possible for one knot to have a steeper spectrum,
but not all.) This places further constraints on the emission models, such that a simple 
power-law energy distribution of a single population of electrons is not viable. We now
discuss models in which the X-ray emission is either synchrotron radiation or
upscattered CMB photons.

\subsection{Synchrotron Model}

The SEDs of $C2$ and $C6$ are each consistent with synchrotron
emission from a single population of relativistic electrons. However, either the slopes of the
electron energy distributions of the two knots are different or the SED contains a hump at
X-ray frequencies owing to inefficient inverse Compton cooling of CMB photons in
the Klein-Nishina regime \citep{DA02,AD04}. Because the optical upper limit falls below the
extrapolation of the radio spectrum, the SED of knot $C4$ can be fit by synchrotron
emission from a single electron population only under the \citet{DA02} scheme. In order
to fit SEDs similar to those of knots $C2$ and $C4$, one needs to
adopt a very high value ($\gtrsim 1000$) of the ``minimum'' electron Lorentz factor
$\gamma_{\rm min}$. This can represent either a lower-energy cut-off or the energy above which
the power-law energy distribution breaks to a slope steeper than $-2$. On the other hand,
relativistic shock waves should heat electron-proton plasmas to an average thermal energy
per particle $\sim \Gamma_{\rm s}' m_{\rm p}c^2$ \citep{BM77}, where $m_{\rm p}$ is the rest
mass of a proton and $\Gamma_{\rm s}'$ is the bulk Lorentz factor of the shock front in the
co-moving frame of the unshocked jet. Hence, $\gamma_{\rm min} \sim 2000 \Gamma_{\rm s}'$ is
expected in the co-moving frame of the emitting plasma.

The main difficulty in applying the \citet{DA02} model to 0827+243 is the poor match at the
low-frequency end of the SED. Knot $C4$ has a steeper radio spectral index than that
of the model SEDs. In addition, the low radio flux of knot $C2$ is expected only
for very short times since the onset of particle acceleration, $\lesssim 10^3$ yr as
measured in the co-moving frame. On the other hand, the latter condition can be met if the
bulk Lorentz factor $\Gamma \gtrsim 20$, since the measured diameter of $C2$ of 5 kpc
implies a lifetime of at least
$2\times 10^4 \Gamma^{-1}$ yr in the co-moving frame.  The SEDs of the knots of
0827+243 then require $\gamma_{\rm min} \gtrsim 10^3$ in order to avoid the dominance of
IC/CMB emission over synchrotron radiation at X-ray energies.

Alternatively, all of the observed SEDs could be explained through synchrotron radiation from
two populations of relativistic electrons, plausibly associated with jet structure transverse to
the axis, as in spine-sheath models \citep[e.g.][]{L99}.
The physical parameters of the plasma corresponding to these two populations are given in
Table 3. They are computed with eq. (A3) under the assumptions that each population is
in energy equipartition with the magnetic field ($A=1$), the kinetic energy in protons and
electrons is comparable ($k=1$), and the emission fills the entire structure ($\phi=1$).
The simulated SEDs are shown in Figure 7.
Constraints on the spine population are similar to those of the \cite{DA02} model,
hence $\Gamma \gtrsim 20$ is needed and the X-ray spectrum is flattened
because the dominant process by which the electrons cool is IC/CMB scattering in the
Klein-Nishina limit. In this case, the energy loss rate depends on electron Lorentz factor as
$\sim \gamma^{0.3}$ \citep{DA02}.
Therefore, the spine is populated by high-energy electrons with 
Lorentz factors $10^3\leq\gamma\leq 3\times 10^8$, flat spectral energy distribution,
$\alpha\sim 0.4$, Doppler factor $\delta \sim 20$,
and low magnetic field, $B\sim 0.5~\mu$G. This value of $\gamma_{\rm max}$ is the minimum
capable of producing the observed X-ray flux via the synchrotron mechanism given the
derived strength of the magnetic field.

The sheath in this model has $\gamma_{\rm max}\sim 5\times 10^5$ and a steep spectral index,
$\alpha\sim 1.5$, and is less relativistic ($\delta\sim 2-5$, although this is not well
specified in the model) with a higher magnetic field, $B\sim$ 20-100 $\mu$G.
The range of possible values of $\gamma_{\rm max}$ allows the synchrotron
power-law spectrum to extend to 15 GHz but not to 420 THz in the optical region. 
The value of the minimum electron Lorentz factor is uncertain;
that given in Table 3 fits the SEDs without excessive energy requirements.
Kelvin-Helmholtz instabilities should generate turbulence
between the spine and sheath, leading to cooling and deceleration of the spine
and an increase in the internal energy of the sheath. We expect that this could
accelerate electrons to achieve a
flatter spectrum in the sheath as the turbulence develops along the jet.
In this model the spine produces the X-rays and the sheath generates the radio emission that we
observe. Since the X-ray and radio
emission regions are not co-spatial, the features with detected X-ray and radio emission might
appear shifted relative to each other. An apparent offset of $\sim 0.4''$ is in fact
observed in the case of knot $C3$ (see Fig. 4), which is especially interesting if real
because the shift is primarily perpendicular to the jet axis.

\subsection{Inverse Compton Scattering off the Cosmic Microwave Background}

Inverse Compton scattering off CMB photons is a potentially efficient way to produce
extended X-ray emission \citep{Tav00, Cel01}. If the mechanism dominates in a kiloparsec-scale
jet, as deduced in a number of objects by \citet{SAM04}, the high-energy photon spectrum
is expected to extend up to the $\gamma$-ray domain. As is discussed by the above authors
(see also eq. A5 in Appendix A), IC/CMB can dominate the X-ray emission if the Doppler factor
and redshift are high. In the case of the quasar 0827+243, a number of properties of the
jet indicate that the Doppler factor remains high out to the X-ray knots:
(1) the X-ray jet aligns with the parsec-scale jet, where highly superluminal apparent speed,
$\beta_{app}=(22\pm 2)$c, is found \citep{J01}; ii) the sharp bend of the X-ray jet is
almost surely amplified owing to a small angle between the jet axis and line of
sight (see above); iii) the decrease of X-ray to radio intensity
ratio along the jet can be attributed to deceleration of the jet from an initially
very high bulk Lorentz factor $\Gamma$ \citep[see][]{GK04}.

Figure 8 shows the observed and simulated SEDs for components $C2, C4$, and $C6$.
The simulated SEDs are combined synchrotron and IC/CMB emission from
a single population of relativistic electrons with $\gamma_{min}=15$ and
$\gamma_{max}=5\times 10^5$. The value of $\gamma_{max}$ should not be higher and
$\gamma_{min}$ should not be lower than the indicated values, otherwise we would expect
these knots to be detected at optical wavelengths. The value of $\gamma_{min}$ cannot
be grater than 15 without causing the low-frequency cutoff to the X-ray spectrum 
to exceed the soft band of our {\it Chandra} observation. 
Because we have no measurements of
$\alpha_x$ and $\alpha_r$ in the same feature
(knots $C3$ and $C4$ are too faint to estimate the X-ray spectral index separately
from $C1$ and $C2$), we consider that $\alpha_x=\alpha_r$,
as expected for IC/CMB emission when the same power-law distribution of scattering
electrons also produces the radio emission. We could also obtain $\alpha_x$ flatter than
$\alpha_r$ if we were to assume that the electron energy distribution breaks from
a slope of $-1.8$ below $\gamma \sim 2000$ to a slope steeper than $-3$ at higher energies;
see the previous section.

The IC/CMB X-ray model combined with synchrotron radio emission allows
us to estimate the parameters of the jet using eq. (A7), (A5), and
(A3).  Equation (A7) provides an estimate for (or limit to) the ratio $\delta B^{-1}$,
while eq. (A3) and (A5) give the values of (or limits to) the products $B\delta$ and
$B\delta^{\alpha+2}$ for the radio synchrotron and X-ray IC/CMB emission, respectively.
Table 4 presents the parameters derived for each jet feature. The model fits
the observed spectral energy distribution quite well (Fig. 8) except that the X-ray
spectrum of knot $C4$ is much steeper than the combined spectrum of all the
X-ray knots. This can be accommodated by the data, since $C4$ contains only about 18\% of
the X-ray flux from the extended jet. It is also consistent with low-energy flattening of the
electron energy distribution, as discussed above. Knot $C2$ has marginally detected
radio emission at 5~GHz (Fig. 4) that matches the flux of the model SED.

One striking property of the model is that the Doppler factor $\delta$ of X-ray features
$B2, C1$, and $C2$ coincides with the estimate of the parsec-scale $\delta$
based on the apparent speed. Moreover, since the X-ray and parsec-scale radio jets
lie in the same direction, there is no evidence of bending that would change the
angle $\theta \lesssim 2^\circ\kern -0.35em .5$ between the jet axis and the line of sight.
This strongly implies that the bulk Lorentz factor remains constant from parsec to
kiloparsec scales and that the magnetic field is very low, $B\approx 2~{\mu}G$,
upstream of the bend. These conclusions would be modified somewhat if a dual population
of electrons were to exist in a given knot such that the X-rays were a roughly
equal combination of synchrotron and IC/CMB emission. In this case, $\delta \sim 7.5$ and
$B \sim 3$ $\mu$G would be possible before the bend. It is doubtful, however, that such a
balance between the two emission mechanisms would occur for all three knots upstream
of the bend.

According to the model parameters, deceleration of the jet does occur after the sharp bend
beyond component $C2$, as anticipated from hydrodynamic considerations. The decrease in
bulk Lorentz factor is accompanied by intensification of the magnetic field. As has been
discussed by \citet{GK04}, this is expected for a decelerating relativistic jet, since
the magnetic field in the co-moving frame of the emitting plasma $B~\propto B_*/\Gamma$, where
$B_*$ is the magnetic field in the rest frame of the host galaxy of the quasar.
We note that these inferences are
independent of the nature of the X-ray emission, but do assume that the equipartition
condition holds. The significant change in the projected direction of the jet implies that the
variation in $\delta$ is influenced both by a decrease in $\Gamma$ and a change in
$\theta$. The apparent bending requires a change $\Delta\theta \ge \theta$, but it would
be difficult to deflect a powerful, highly relativistic jet by more than a few degrees
while conserving momentum. Hence, we can assume that the new viewing angle
$\theta$(post-bend) $\approx 5^\circ$. As shown in Figure 9, the
value of $\Gamma$ after the bend is then roughly equal to half the Doppler factor.
Therefore, the decrease in the Doppler factor is primarily a consequence of deceleration of
the jet flow.

Further evidence for deceleration of the jet flow is found in the shift
between the X-ray and radio positions of component $C3$ and in the existence
of an extended radio tail downstream of knot $C4$ (see Fig. 4), in qualitative 
agreement with numerical simulations of a decelerating jet (cf. Fig. 2 in \citealt{GK04}).

The required kinetic luminosities of the X-ray emitting knots are high, but perhaps
possible for a
powerful jet, $\sim 7\times 10^{47}$ erg s$^{-1}$ for $C3$, $\sim 6\times 10^{47}$ erg s$^{-1}$
for $C4$, and less for the others, when the bulk kinetic energy of the protons is included.
These values argue that the jet cannot be far from equipartition between the energy density
of relativistic electrons and that of the magnetic field. Otherwise, the required luminosities
would be excessive \citep[see][]{AD04,DA04}.
If the positively charged particles are mainly positrons, the luminosities are quite
reasonable, $\lesssim 10^{46}$ erg s$^{-1}$, and some departure from equipartition would be
possible. The required kinetic luminosities of knots $C3$ and $C4$ could also be reduced
below $3\times 10^{46}$ erg s$^{-1}$ if there is a break in the electron energy distribution
at $\gamma_{\rm break}\sim 2000$, in which case $\gamma_{\rm min}$ in eq. (A3) should be
replaced by $\gamma_{\rm break}$. As mentioned above, this can explain the flat X-ray
spectra alongside the steeper radio spectra.

\section{Conclusions}
The relatively flat X-ray spectrum of the extended jet of 0827+243 is consistent with
synchrotron radiation from extremely high-energy electrons under the \citet{DA02} scenario
or within a dual population (e.g., spine-sheath) model. It can also be explained by inverse 
Compton scattering of CMB photons by $\gamma \sim 15$ electrons if the energy distribution
of the electrons is relatively flat below
$\gamma \sim 10^3$ and steeper above this value. All models
require very high Doppler factors, consistent with the value $\delta \approx 20$ inferred from
the parsec-scale superluminal apparent motion, and very low magnetic field ($\lesssim 2~{\mu}G$)
upstream of the bend situated 5$''$ from the core. The implication is that the bulk Lorentz
factor of the jet flow maintains its high value all the way from parsec scales to $\sim 800$
kpc from the nucleus. Beyond this point, the jet bends by about $90^\circ$ projected on the
sky and at least $2^\circ\kern -0.35em .4$ in three dimensions. In the IC/CMB model,
the flow decelerates while the magnetic field intensifies at and beyond the bend. The latter
follows the expectations of the model of \citet{GK04}, although the jump in field strength
at the bend is about 4 times greater than the model predicts. This is in line with the
supposition that a standing shock accompanies the bend in the flow.

In the IC/CMB case and in most of the synchrotron models, the energetics of the
jet are reasonable if the energy distribution of the electrons is
less steep than $\propto \gamma^{-2}$ below a rather high value of $\gamma$, $\sim 1000$.
Such high break energies are consistent with particle acceleration by relativistic shocks.
Other possible processes for producing high values of $\gamma_{\rm min}$ are discussed
by \citet{GBW04}.
As mentioned by \citet{GK04}, the \citet{DA02} proposal predicts that the sizes of X-ray
knots should be smaller than those of optical knots owing to the more sever energy losses of
X-ray emitting electrons. However, we detect no optical knots in 0827+243, and hence cannot
apply this test with our current data.

The flat spectrum of the X-ray jet features suggests that the maximum
energy output of the knots could occur in the $\gamma$-ray domain, perhaps up to
the TeV region ($\nu\sim 10^{25}$~Hz), although this emission is difficult to observe.
TeV photons from such a high redshift ($z=0.939$) will pair-produce
off cosmic infrared photons before reaching the Earth. In
addition, current $\gamma$-ray detectors have insufficient resolution to isolate
any emission from the extended jet, and the nuclear emission probably dominates the
total $\gamma$-ray flux. Nevertheless, the $\gamma$-ray flux should not be observed to drop
below the expected level if the models are valid.
At lower frequencies, synchrotron and, for knots with low values
of $\delta$, IC/CMB emission could contribute substantial amounts of
optical and infrared flux from the jet in 0827+243. Observations in these wavebands
could define the lower and/or upper cutoffs of the relativistic electron energy distributions,
thereby removing one of the remaining ambiguities in our analysis.

\appendix
\section{Derivation of Magnetic Field and Doppler Factor from
X-ray and Radio Observations}

\subsection{Relations for Synchrotron Emission}

If the dominant contribution to the observed flux density at a given frequency can
be identified as synchrotron radiation, we can relate the magnetic field strength $B$
and Doppler factor $\delta$ to observed parameters as follows. Based on the
formulas given in \citet{P70}, \citet{M83} expressed the synchrotron flux
density $F_\nu$ at frequency $\nu_{\rm s}$ as
\begin{equation}
F_{\rm s}(\nu_{\rm s}) = c_1(\alpha) D_\ell^{-2} N_0 B^{1+\alpha} V
\delta^{3+\alpha} (1+z)^{1-\alpha} \nu_{\rm s}^{-\alpha},
\end{equation}
where $c_1(\alpha)$ is tabulated in Table A1, $D_\ell$ is the luminosity distance, and
$N_0$ and $\alpha$ define the electron energy distribution,
$N(E)=N_0E^{-(1+2\alpha)}$. The volume of the emitting region $V =(4\pi/3)R^3\phi$
if the source is approximated as a sphere with a volume filling factor $\phi$.

The quantity $N_0$ can be related to the magnetic field $B$ through the ratio of
particle to magnetic energy density $A$ and the fraction $k$
of the particle energy contained in non-electrons (where positrons are considered as
electrons). Equipartition in energy between particles and magnetic field corresponds
to $A = 1$. By integrating over $N(E)$ to determine
the energy density in electrons and relating this
to the magnetic energy density $B^2/(8\pi)$, we find that
\begin{equation}
N_0 = {\frac{A(mc^2)^{2\alpha - 1}}{8\pi(1+k)}}
g(\alpha,\gamma_{\rm min},\gamma_{\rm max})^{-1} B^2,
\end{equation}
where $g(\alpha,\gamma_{\rm min},\gamma_{\rm max}) \equiv (2\alpha - 1)^{-1}
[\gamma_{\rm min}^{1-2\alpha} - \gamma_{\rm max}^{1-2\alpha}]$ except for the special
case $\alpha=0.5$, for which $g = \ln(\gamma_{\rm max}/\gamma_{\rm min})$.

We can express the radius $R$ in terms of the
angular diameter $\Theta_{\rm as}$ arcsec, redshift $z$, and luminosity distance
$D_{\rm Gpc}$ Gpc, as $R = 2.42\Theta_{\rm as} D_{\rm Gpc} (1+z)^{-2}$ kpc.
Inserting the above relations into (A1) and converting to
convenient units as denoted by subscripts, we obtain the value of the magnetic field
that corresponds to the observed flux density of synchrotron emission:
\begin{equation}
B = c_4(\alpha) \delta^{-1} (1+z)^{(5+\alpha)/(3+\alpha)}
[\frac{(1+k)}{A\phi} g(\alpha,\gamma_{\rm min},\gamma_{\rm max})
D_{\rm Gpc}^{-1} \Theta_{\rm as}^{-3} F_{\rm s,mJy} \nu_{\rm s,GHz}^\alpha]^{1/(3+\alpha)}
~~\mu{\rm G}.
\end{equation}
The function $c_4(\alpha)$ is tabulated in Table A1.

\subsection{Combined Inverse Compton and Synchrotron Relations}

The flux density of inverse Compton emission at frequency $\nu_{\rm IC}$
from a population of electrons with a
power-law energy distribution scattering a monochromatic field of seed photons of
frequency $\nu_{\rm seed}$ and energy density $u_{\rm seed}$ is
\begin{equation}
F_{\rm IC} = c_2(\alpha)
{\frac{u_{\rm seed}}{\nu_{\rm seed}}} D_\ell^{-2} N_0 V \delta^{4+2\alpha}
(1+z)^{1-\alpha} (\nu_{\rm IC}/\nu_{\rm seed})^{-\alpha}.
\end{equation}
Here, $c_2(\alpha) \equiv (c\sigma_{\rm T}/2)(mc^2)^{-2\alpha}$, where $c$ is the speed of
light, $m$ is the mass of the electron, and $\sigma_{\rm T}$
is the Thomson cross-section. In order to derive equation (A5), we started with the
inverse Compton formula of \citet{D95} and applied relation (A2).

In the case of IC/CMB emission, $u_{\rm seed} =
4.2\times 10^{-13} (1+z)^4$ erg cm$^{-3}$ and $\nu_{\rm seed} = 5.7\times 10^{10} (1+z)$ Hz.
Use of these values in eq. (A4) and conversion to convenient units gives
\begin{equation}
F_{\rm IC/CMB} =
{\frac{c_5(\alpha) A \phi}{1+k}} g(\alpha,\gamma_{\rm min},\gamma_{\rm max})^{-1}
D_{\rm Gpc} \Theta_{\rm as}^3 B_{\rm \mu G}^2 \delta^{4+2\alpha}
(1+z)^{-2} E_{\rm IC,keV}^{-\alpha}~~{\rm nJy},
\end{equation}
where $B_{\rm \mu G}$ is the magnetic field in $\mu$G and $E_{\rm IC,keV}$ is the
observed scattered photon energy in keV. The function $c_5(\alpha) \equiv
0.063(4.2\times 10^6)^{-\alpha} f(\alpha)$, where $f(\alpha)$ is a
factor that corrects for the use of a monochromatic seed photon field rather than
a Planck function in the derivation of eq. (A5). We tabulate both $c_5$ and $f$, the latter of
which we have calculated through numerical integration, in Table A1. Note that the inverse
proportionality between the IC/CMB flux density and the redshift term is due to the
conversion of linear size to angular size. For a fixed volume of the emission region,
the IC/CMB flux density varies as $(1+z)^4$.

We can substitute expression (A3) for $B_{\rm \mu G}$ into equation (A5) and solve for the
Doppler factor in terms of observable quantities and the function
$g(\alpha,\gamma_{\rm min},\gamma_{\rm max})$:
\begin{eqnarray}
\delta = 10 c_6(\alpha)[\frac{(1+k)}{A \phi}g(\alpha,\gamma_{\rm min},\gamma_{\rm max})
D_{\rm Gpc}^{-1} \Theta_{\rm as}^{-3}]^{1/[2(3+\alpha)]}
[F_{\rm s,mJy} \nu_{\rm s,GHz}^{\alpha} (1+z)^2]^{-1/[(1+\alpha)(3+\alpha)]}\times \nonumber\\
\times [F_{\rm IC/CMB,nJy}(\nu_{\rm IC}) E_{\rm IC,keV}^{\alpha}]^{1/[2(1+\alpha)]}.\ \ \ \
\end{eqnarray}
The function $c_6(\alpha)$, which is of order unity, is listed in Table A1. The exponent
$1/[2(3+\alpha)]$ ranges from 0.16 to 0.11 for the values of $\alpha$ considered
here (0.2 to 1.5), hence the dependence of the derived Doppler factor on the poorly
known quantity $g(\alpha,\gamma_{\rm min},\gamma_{\rm max})$ is quite weak.

If the dominant X-ray emission mechanism of a knot is IC/CMB, we can combine equations
(A1) and (A5) to form the ratio of the
X-ray IC/CMB flux density at photon energy $E_{\rm keV}$ keV to the synchrotron flux density
at frequency $\nu_{\rm s,GHz}$ GHz. This produces a formula involving
$B$ and $\delta$ that is independent of the source geometry and electron energy
distribution, except that the latter
needs to extend over the range corresponding to electrons that radiate synchrotron emission
at frequency $\nu_{\rm s}$ and IC/CMB emission at energy $E_{\rm IC}$:
\begin{equation}
{\frac{F_{\rm IC/CMB}(\nu_{\rm IC})}{F_{\rm s}(\nu_{\rm s})}} = {\frac{c_2(\alpha)}{c_1(\alpha)}}
{\frac{u_{\rm seed}}{\nu_{\rm seed}}} B^{-(1+\alpha)} \delta^{1+\alpha}
\nu_{\rm seed}^\alpha \nu_{\rm IC}^{-\alpha} \nu_{\rm s}^\alpha.
\end{equation}
Expressing this in more convenient units and solving for the magnetic field, we obtain
\begin{equation}
B = c_3(\alpha) \delta (1+z)^{(3+\alpha)/(1+\alpha)}
\left(\frac{F_{\rm s,mJy}\nu_{\rm s,GHz}^{\alpha}}
{F_{\rm IC/CMB,nJy}E_{\rm IC,keV}^{\alpha}}\right)^{1/(1+\alpha)}~~\mu{\rm G}.
\end{equation}

\acknowledgments

The authors thank Dr. D. Harris for advice regarding reduction of {\it Chandra} data
and Dr. C. Dermer for comments on a preliminary draft.
This material is based upon
work supported by the National Aeronautics and Space Administration under
Chandra Guest Investigator grant no. GO2-3137X administered
by the Smithsonian Astrophysical Observatory, and by the U.S.
National Science Foundation under grant no. AST-0098579. The
VLA is a facility of the National Radio Astronomy Observatory,
operated by Associated Universities Inc. under cooperative agreement
with the National Science Foundation.

\begin{deluxetable}{cccc}
\singlespace
\tablecolumns{4}
\tablecaption{\small\bf X-Ray Measurements}
\tabletypesize{\footnotesize}
\tablehead{
\colhead{}&\colhead{Total Counts}&\colhead{Net Count Rate (phot s$^{-1}$)}&\colhead{$HR^a$}
}
\startdata
Core&6315$\pm$79&0.338$\pm$0.004&$-$0.42$\pm$0.02  \\
C1&28$\pm$6&0.0015$\pm$0.0003&$-$0.43$\pm$0.24 \\
C2&30$\pm$6&0.0016$\pm$0.0003&$-$0.31$\pm$0.21  \\
C3&25$\pm$5&0.0013$\pm$0.0003&$-$0.76$\pm$0.25  \\
C4&23$\pm$4&0.0012$\pm$0.0003&$-$0.54$\pm$0.27  \\
Bkg&5$\pm$2&0.0002$\pm$0.0001&  \\
\enddata
\tablecomments{$^a$Hardness ratio, $HR=(H-S)/(H+S),$ where $S$ = net photon counts at 0.2-2~keV
and $H$ = net counts at 2-8~keV.}
\end{deluxetable}
\begin{deluxetable}{cccccccccc}
\singlespace
\tablecaption{\small\bf Parameters of Jet Features}
\tabletypesize{\footnotesize}
\tablehead{
\colhead{}&\multicolumn{4}{c}{\bf X-Rays}&\multicolumn{5}{c}{\bf Radio} \\
\colhead{}&\colhead{R$^a$($''$)}&\colhead{$\Phi^b$($^\circ$)}&
\colhead{$a^c$($''$)}&\colhead{$S_{\rm 1~keV}$ (nJy)}&
\colhead{R($''$)}&\colhead{$\Phi$($^\circ$)}&
\colhead{a($''$)}&\colhead{$S_{\rm 5~GHz}$ (mJy)}&\colhead{$\alpha_{\rm r}^d$}
}
\startdata
Core&0.0$\pm$0.1&0.0&0.80&6185$\pm$3&0.0$\pm$0.05&0.0&0.0&1359$\pm$10&$-$0.39$\pm$0.02 \\
B1$^e$&$\sim$1.0&$\sim$140&\nodata&\nodata&0.5$\pm$0.2&140$\pm$1&0.25$\pm$0.05&4.06$\pm$0.08&0.72$\pm$0.06 \\
B2&2.8&146&1.5&0.66$\pm$0.21&\nodata&\nodata&\nodata&$<$0.06&\nodata \\
C1&4.1&150&0.7&0.77$\pm$0.21&\nodata&\nodata&\nodata&$<$0.06&\nodata \\
C2&4.9&151&0.7&0.91$\pm$0.23&\nodata&\nodata&\nodata&$<$0.06&\nodata  \\
C3&5.8&157&0.9&0.82$\pm$0.21&5.6$\pm$0.2&162$\pm$2&0.80$\pm$0.05&1.2$\pm$0.1&$\geq$1.5  \\
C4&6.2&173&1.2&0.56$\pm$0.19&6.2$\pm$0.2&177$\pm$2&1.20$\pm$0.05&3.5$\pm$0.2&1.13$\pm$0.10  \\
C5&\nodata&\nodata&\nodata&$<$0.10&7.4$\pm$0.2&$-$166$\pm$2&1.50$\pm$0.05&5.4$\pm$0.3&1.13$\pm$0.06 \\
C6&\nodata&\nodata&\nodata&$<$0.10&8.1$\pm$0.2&$-$160$\pm$2&0.70$\pm$0.05&9.5$\pm$0.2&0.84$\pm$0.04 \\
\enddata
\tablecomments{$^a$distance from the core; $^b$position angle of knot relative
to the core; $^c$size of component defined as FWHM of best-fit gaussian; uncertainly in
value is large, $\sim 50$\%;
$^d$spectral index between 5 and 15 GHz defined as $S_\nu\propto\nu^{-\alpha}$; $^e$the X-ray
characteristics of knot $B1$ are not well determined owing to uncertainties in the
PSF subtraction}
\end{deluxetable}
\begin{deluxetable}{cccccccccc}
\singlespace
\tablecolumns{10}
\tablecaption{\small\bf Physical Parameters of the Jet: Synchrotron Model}
\tabletypesize{\footnotesize}
\tablehead{
\colhead{}&\colhead{$R''$}&\multicolumn{4}{c}{Spine Population}&
\multicolumn{4}{c}{Sheath Population} \\
\colhead{Knot}&\colhead{}&\colhead{$\alpha$}&\colhead{$\gamma_{\rm min}$}&\colhead{$\gamma_{\rm max}$}&
\colhead{$B\delta$ ($\mu$G)}&\colhead{$\alpha$}&\colhead{$\gamma_{\rm min}$} &\colhead{$\gamma_{\rm max}$}&
\colhead{$B\delta$ ($\mu$G)}
}
\startdata
B2&2.8& 0.4&$10^3$&3.0$\times 10^8$&5&1.5&$10^2$&5$\times 10^5$&$<60$ \\
C1&4.1&0.4&$10^3$&3.0$\times 10^8$&10&1.5&$10^2$&5$\times 10^5$&$<100$ \\
C2&4.9&0.4&$10^3$&3.0$\times 10^8$&10&1.5&$10^2$&5$\times 10^5$&$<100$ \\
C3&5.8&0.4&$10^3$&2.5$\times 10^8$&10&1.5&$10^2$&5$\times 10^5$&200 \\
C4&6.2&0.4&$10^3$&2.5$\times 10^8$&5&1.1&$10^2$&5$\times 10^5$&100 \\
C5&7.4&0.4&$10^3$&$\leq 1\times 10^7$&$<3$&1.1&$10^2$&5$\times 10^5$&90 \\
C6&8.1&0.4&$10^3$&$\leq 1\times 10^7$&$<4$&0.8&$10^2$&5$\times 10^5$&90 \\
\enddata 
\end{deluxetable}
\begin{deluxetable}{ccccccc}
\singlespace
\tablecolumns{7}
\tablecaption{\small\bf Physical Parameters of the Jet: IC/CMB model}
\tabletypesize{\footnotesize}
\tablehead{
\colhead{Knot}&\colhead{$R('')$}&\colhead{$\alpha$}&\colhead{$B$ ($\mu$G)}&
\colhead{$\delta$}&\colhead{$\Gamma$}&\colhead{$\theta(^\circ)$}
}
\startdata
B2&2.8&0.4&2&24&24&2.4 \\
C1&4.1&0.4&2&24&24&2.4 \\
C2&4.9&0.4&2&24&24&2.4 \\
C3&5.8&1.5&70&4.6&2.5&4.8 \\
C4&6.2&1.1&60&3.3&1.8&4.8 \\
C5&7.4&1.1&100&2.3&1.4&4.8  \\
C6&8.1&0.8&100&2.3&1.4&4.8 \\
\enddata
\end{deluxetable}

\begin{deluxetable}{llllllll}
\singlespace
\tablenum {A1}
\tablecolumns{8}
\tablecaption{\small\bf Functions of Spectral Index $\alpha$}
\tablehead{
\colhead{$\alpha$}&\colhead{$c_1$}&\colhead{$c_2$}&\colhead{$c_3$}&\colhead{$c_4$}&
\colhead{$c_5$}&\colhead{$c_6$}&\colhead{$f$}
}
\startdata
$0.2$&$1.2\times 10^{-19}$&$2.7\times 10^{-12}$&0.018&1.4&$1.3\times 10^{-3}$&1.20&0.44 \\
$0.3$&$8.2\times 10^{-18}$&$4.5\times 10^{-11}$&0.036&2.5&$3.1\times 10^{-4}$&1.10&0.48 \\
$0.4$&$5.4\times 10^{-16}$&$7.4\times 10^{-10}$&0.068&4.4&$7.3\times 10^{-5}$&1.04&0.52 \\
$0.5$&$3.5\times 10^{-14}$&$1.2\times 10^{-8}$&0.12&7.8&$1.8\times 10^{-5}$&0.97&0.58 \\
$0.6$&$2.5\times 10^{-12}$&$2.0\times 10^{-7}$&0.18&12&$4.3\times 10^{-6}$&1.00&0.64 \\
$0.7$&$1.7\times 10^{-10}$&$3.3\times 10^{-6}$&0.26&20&$1.0\times 10^{-6}$&1.00&0.71 \\
$0.8$&$1.2\times 10^{-8}$&$5.5\times 10^{-5}$&0.37&32&$2.5\times 10^{-7}$&1.00&0.79 \\
$0.9$&$8.8\times 10^{-7}$&$9.0\times 10^{-4}$&0.49&48&$6.1\times 10^{-8}$&1.03&0.89 \\
$1.0$&$6.3\times 10^{-5}$&0.015&0.64&74&$1.5\times 10^{-8}$&1.05&1.00 \\
$1.1$&$4.5\times 10^{-3}$&0.25&0.82&100&$3.7\times 10^{-9}$&1.14&1.13 \\
$1.2$&0.33&4.1&1.0&150&$9.1\times 10^{-10}$&1.16&1.28 \\
$1.3$&23&67&1.2&210&$2.2\times 10^{-10}$&1.23&1.45 \\
$1.4$&$1.8\times 10^3$&$1.1\times 10^3$&1.4&290&$5.5\times 10^{-11}$&1.29&1.65 \\
$1.5$&$1.5\times 10^5$&$1.8\times 10^4$&1.6&380&$1.4\times 10^{-11}$&1.38&1.89
\enddata
\end{deluxetable}
\newpage
\begin{figure}
\epsscale{1.0}
\plotone{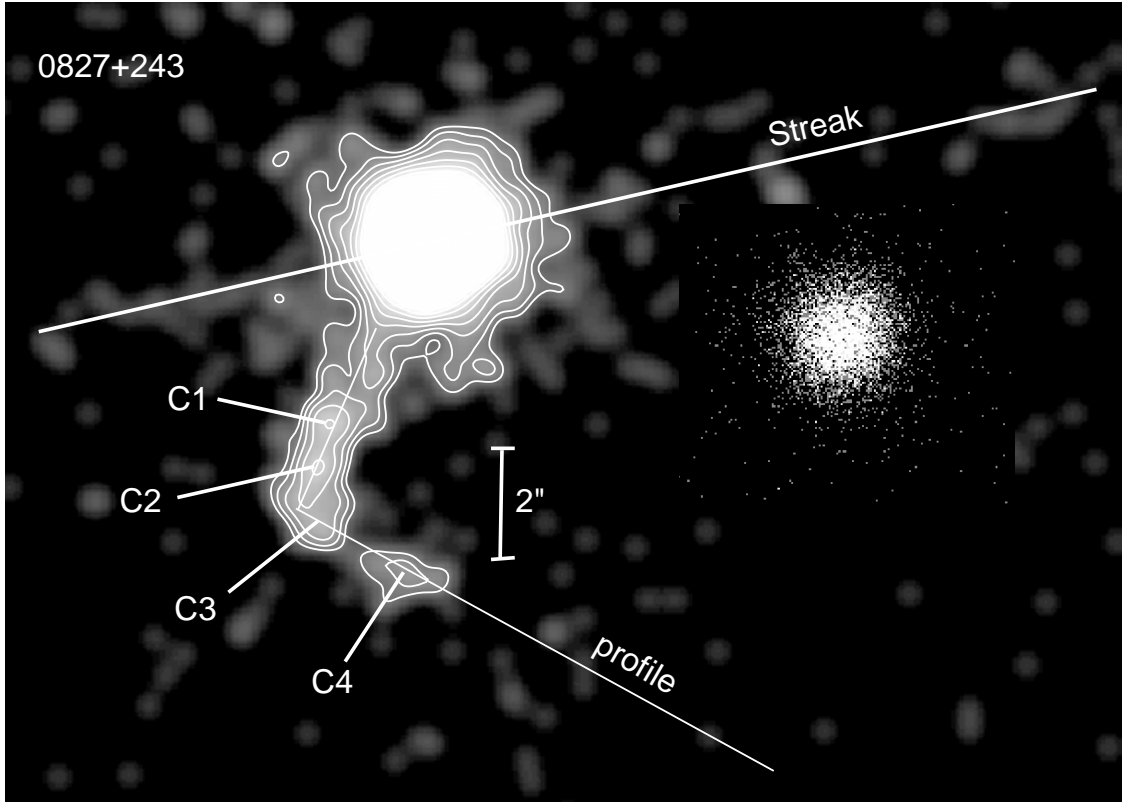}
\caption{ASIC-S3 Chandra subpixelated image of 0827+243 at 0.2-8~keV convolved with
a Gaussian kernel of $\sigma$=$0''\kern -0.35em .5$.  The lines indicate 
positions of the readout streak and jet axis. The contour levels are in factors of 2
from 128 to 2 counts pixel$^{-1}$. The X-ray knots are labeled.
The simulated PSF for the core region is inset on the west side of the image.\label{fig1}}
\end{figure}
\begin{figure}
\plottwo{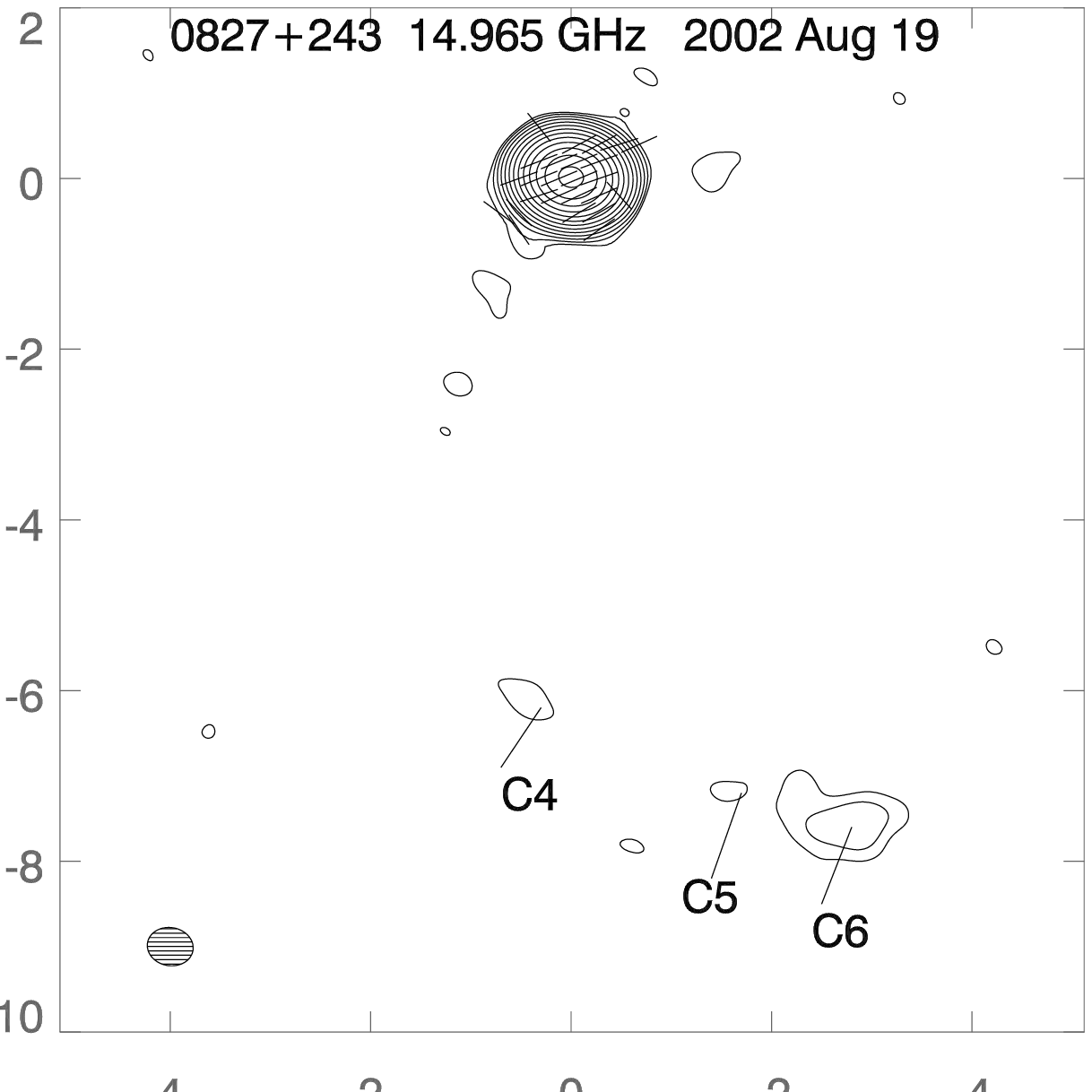}{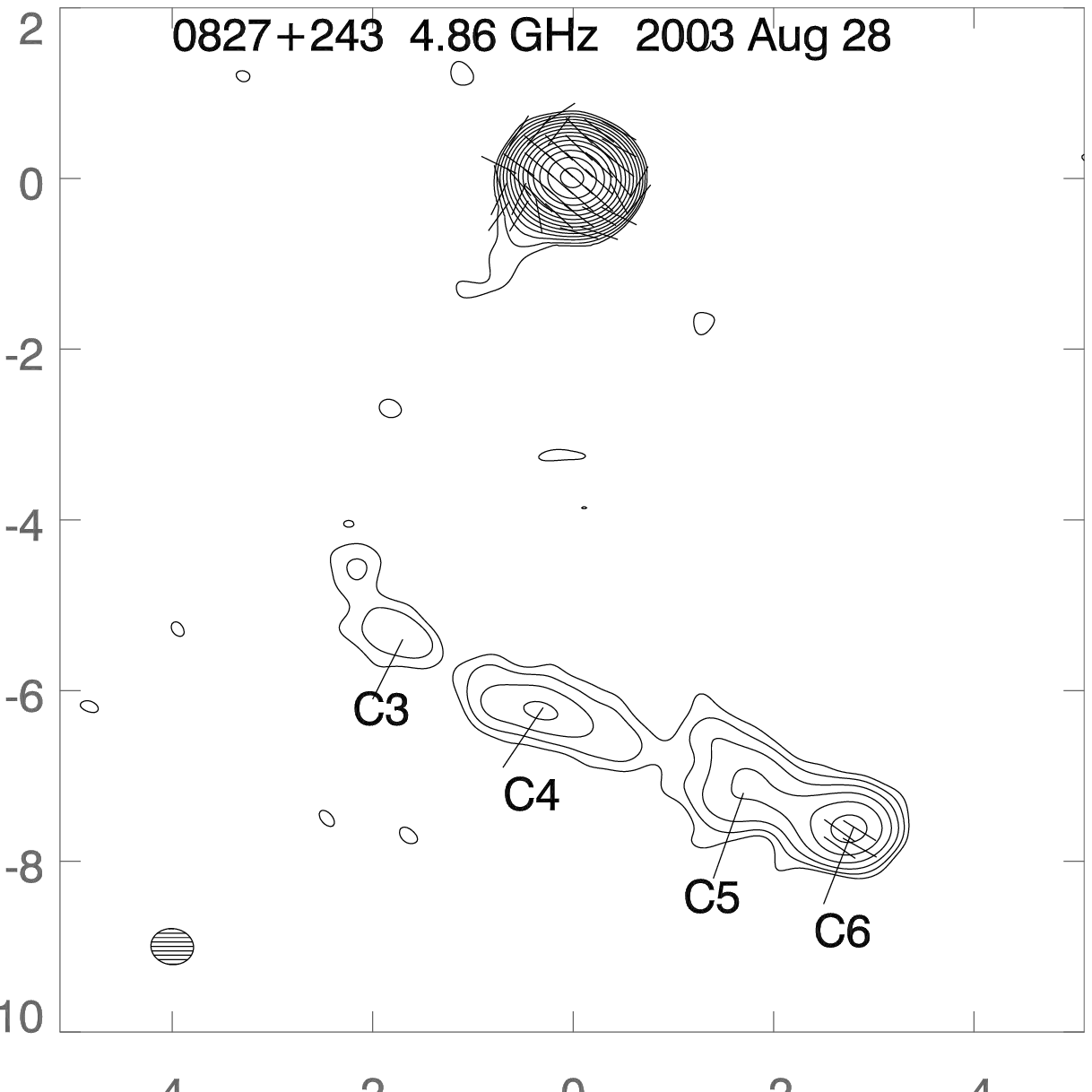}
\caption{VLA images of 0827+243. The sticks indicate
the direction of the electric vectors. At 15~GHz (left panel) the total intensity peak
is 2.09~Jy/beam, the CLEAN beam is 0.468$\times$0.438 arcsec$^2$ at $52^\circ\kern -0.35em .8$,
the RMS is 0.075~mJy/beam, the lowest contour is 0.02\% of the peak and
contours increase by factors of 2. At 4.9~GHz (right panel) the total intensity peak  is
1.36~Jy/beam, the CLEAN beam is 0.432$\times$0.416 arcsec$^2$ at $48^\circ\kern -0.35em .4$,
the RMS is 0.025~mJy/beam, the lowest contour is 0.01\% of the peak and
contours increase by factors of 2. \label{fig2}}
\end{figure} 
\begin{figure}
\plottwo{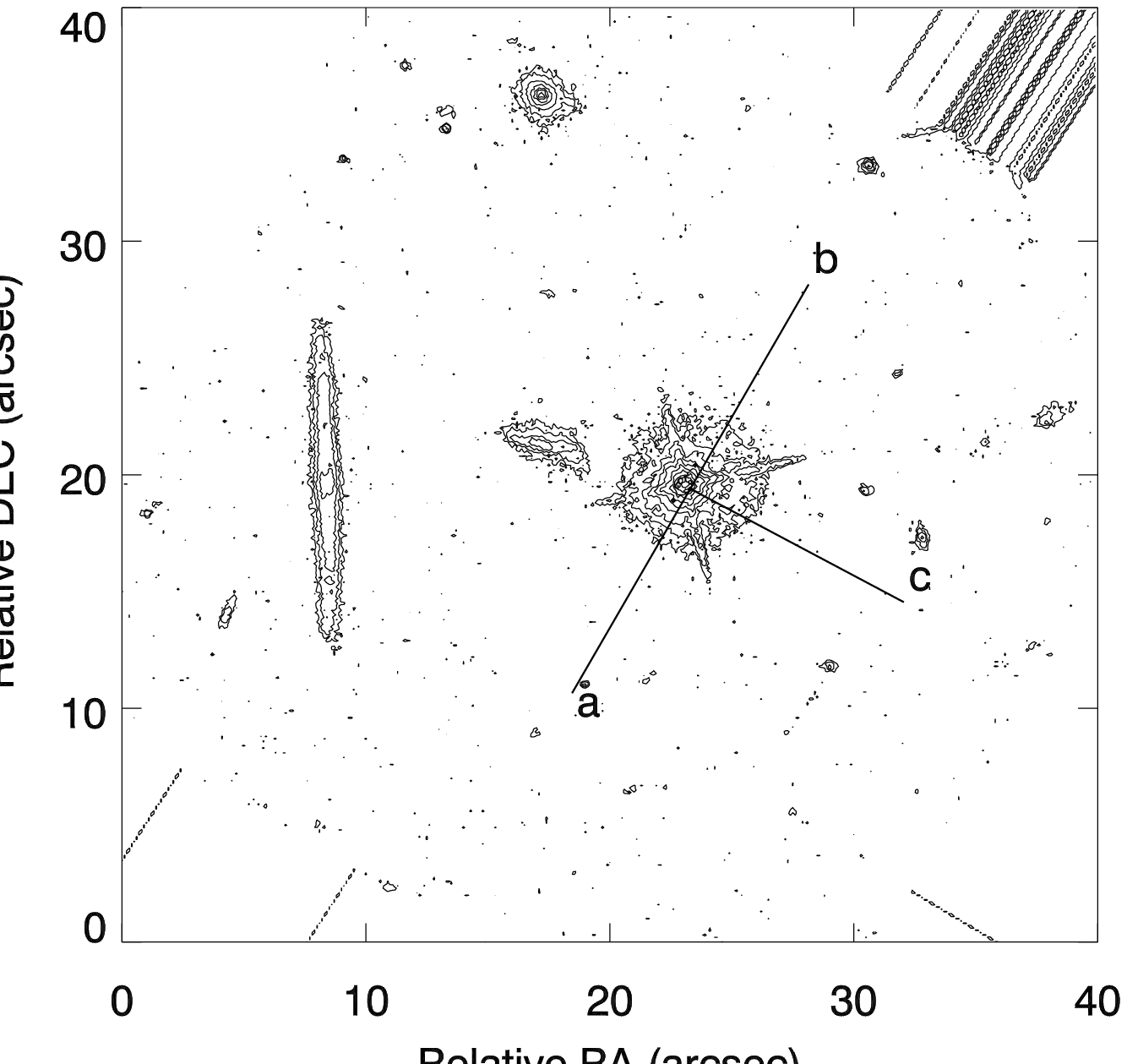}{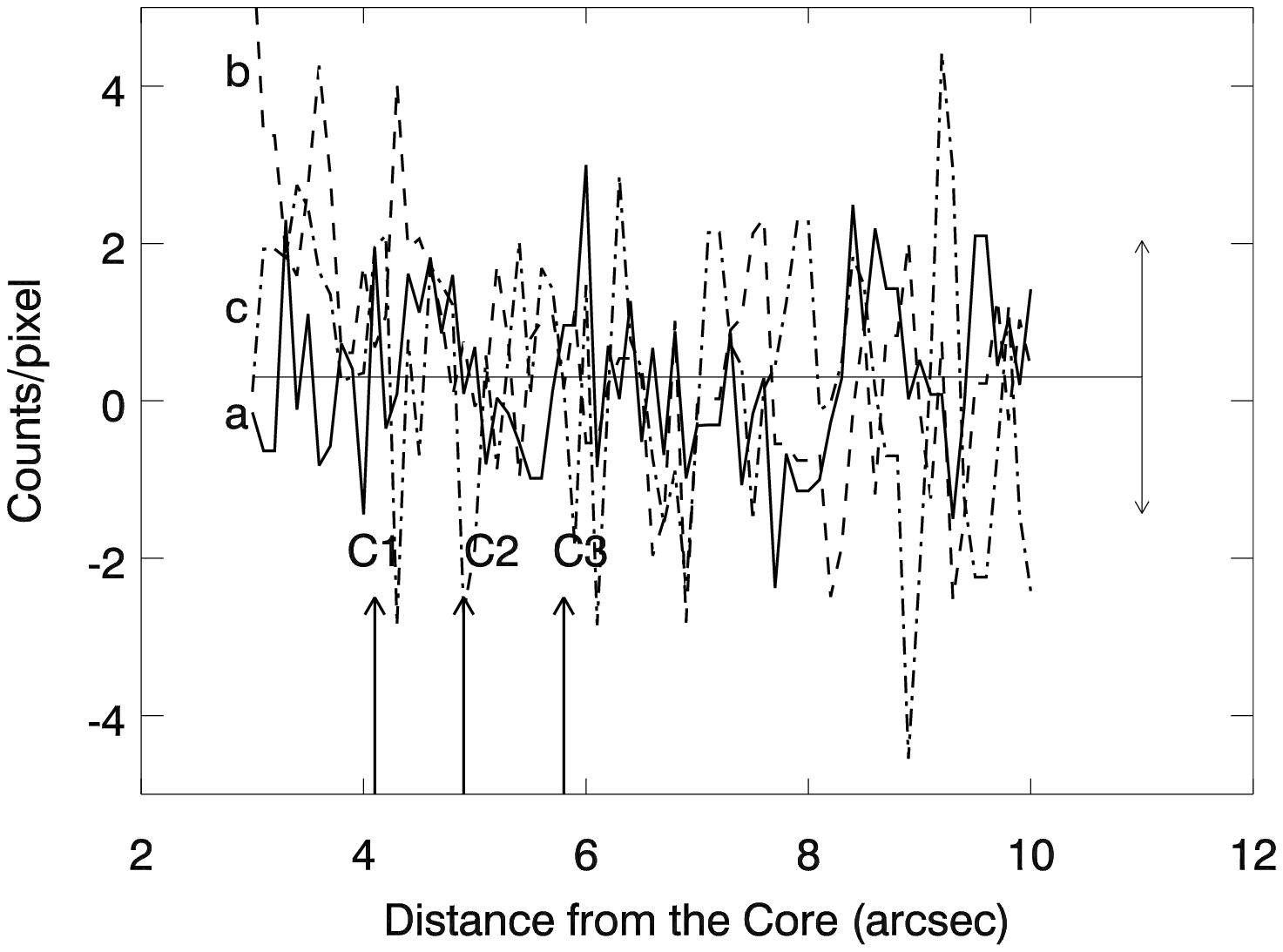}
\caption{{\it Left panel: HST} image of 0827+243 with pixel size $0''\kern -0.35em .1$. The
contours are in factors of 2 from 0.0001 to 0.768 of the optical peak of
3262 counts/pixel. The solid lines $a, b,$ and $c$ indicate positions of the X-ray jet,
counter-jet, and control ray, respectively. {\it Right panel:} the optical profiles along the
X-ray jet $a$ (solid line), counter-jet $b$ (dashed line), and control ray $c$ (dash-dot line).
The arrows show location of the X-ray knots. The solid straight line with arrows indicates the
background level and its uncertainty.
\label{fig4}}
\end{figure} 
\begin{figure}
\epsscale{1.0}
\plotone{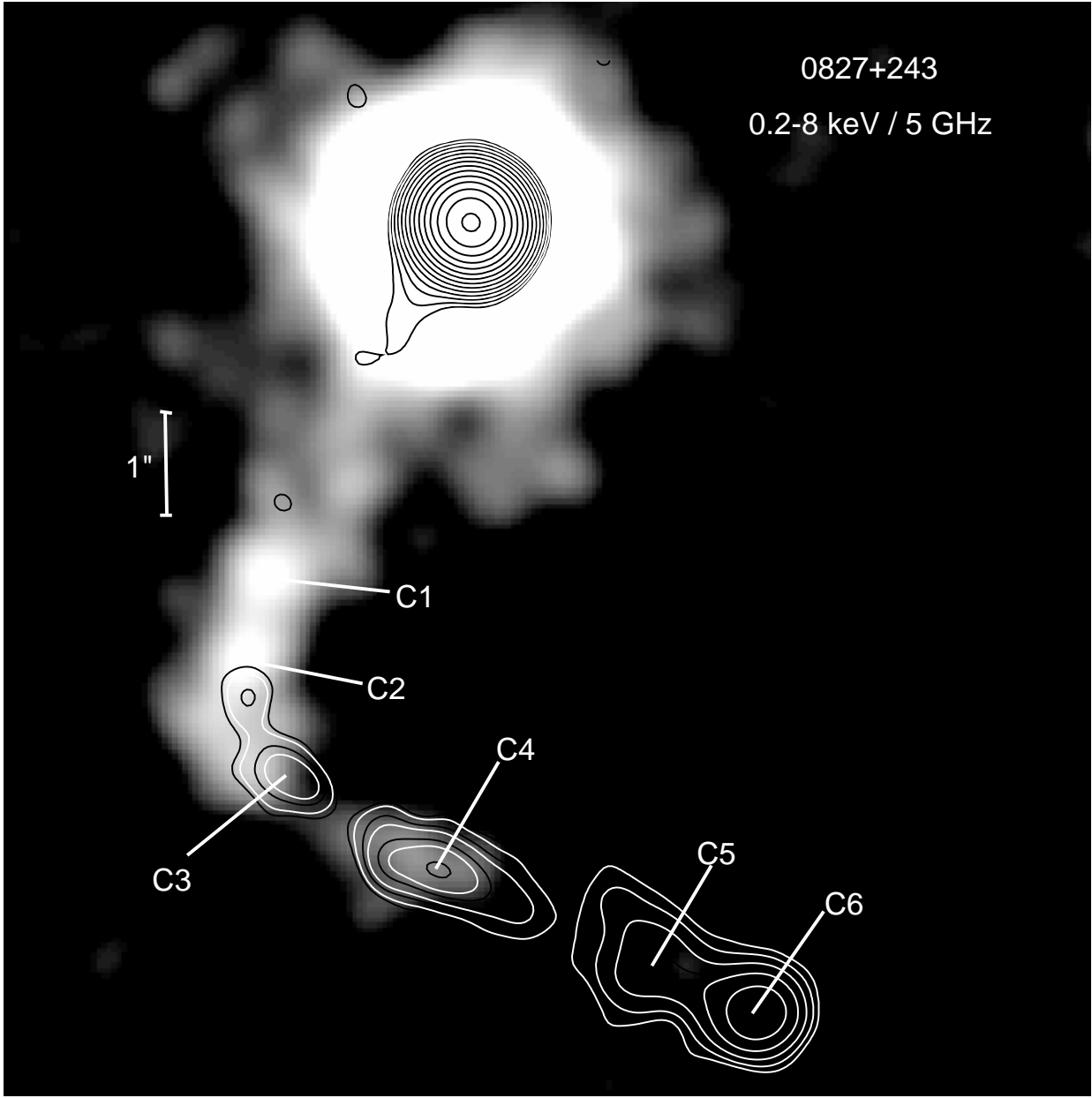}
\caption{ASIC-S3 Chandra image of 0827+243 at 0.2-8~keV (gray scale)
with the 4.9~GHz VLA image (contours) superposed. The images are convolved
with the same Gaussian beam of FWHM $0''\kern -0.35em .5\times 0''\kern -0.35em .5$. \label{fig5}}
\end{figure}
\begin{figure}
\epsscale{1.0}
\plotone{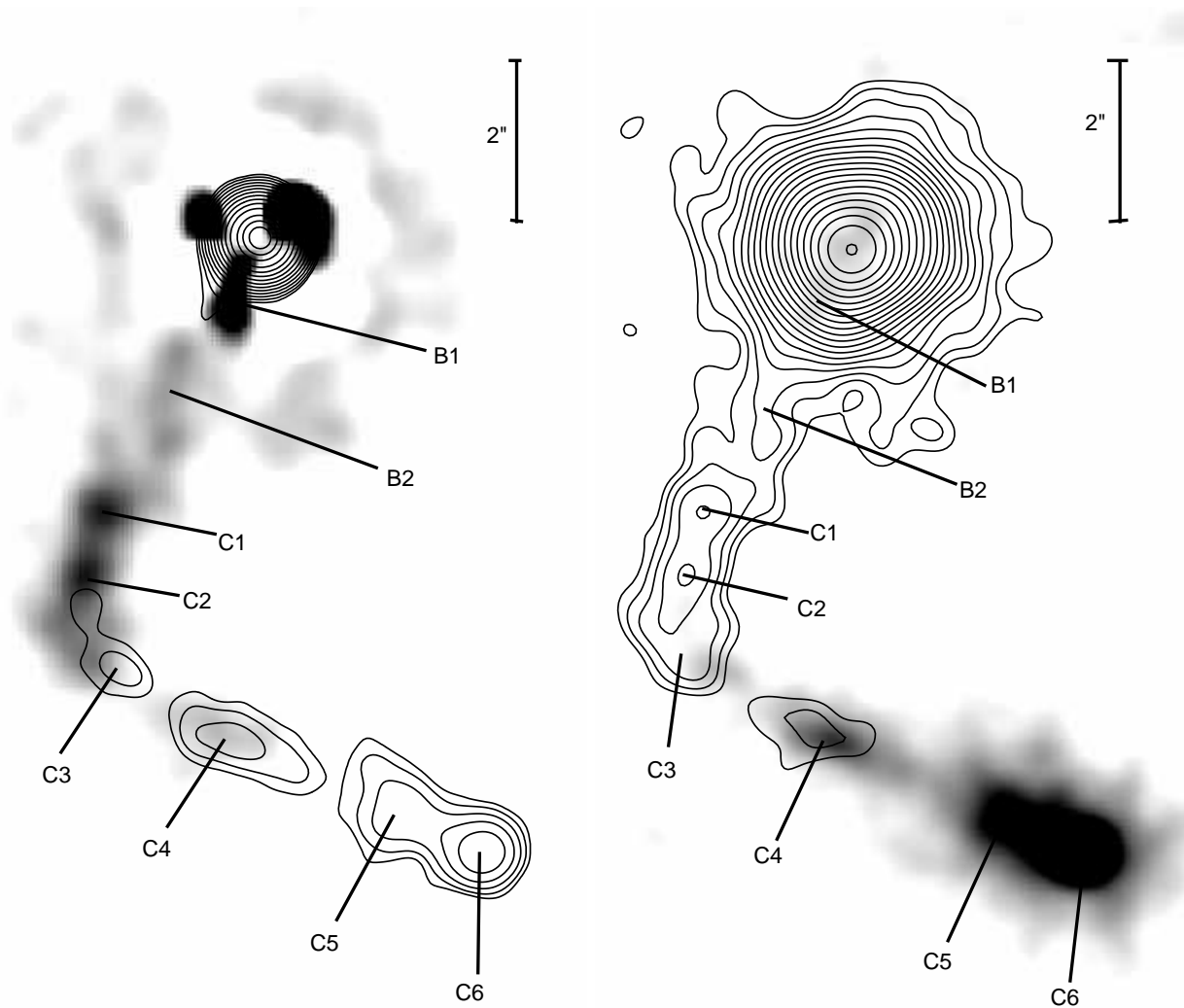}
\caption{{\it Left panel:} ASIC-S3 Chandra image of 0827+243 at 0.2-8~keV (gray scale)
deconvolved by using the PSF shown in Fig.1. The core PSF is subtracted from
the image. Superposed on the image are the 4.9~GHz VLA contours (see Fig. 2).
{\it Right panel:}  VLA image of the quasar at 4.9~GHz
(gray scale). The radio core is subtracted from the image after being modeled with
a circular Gaussian of FWHM=0.01~mas. Superposed on the image are the X-ray
contours. \label{fig6}}
\end{figure}
\begin{figure}
\epsscale{1.0}
\plotone{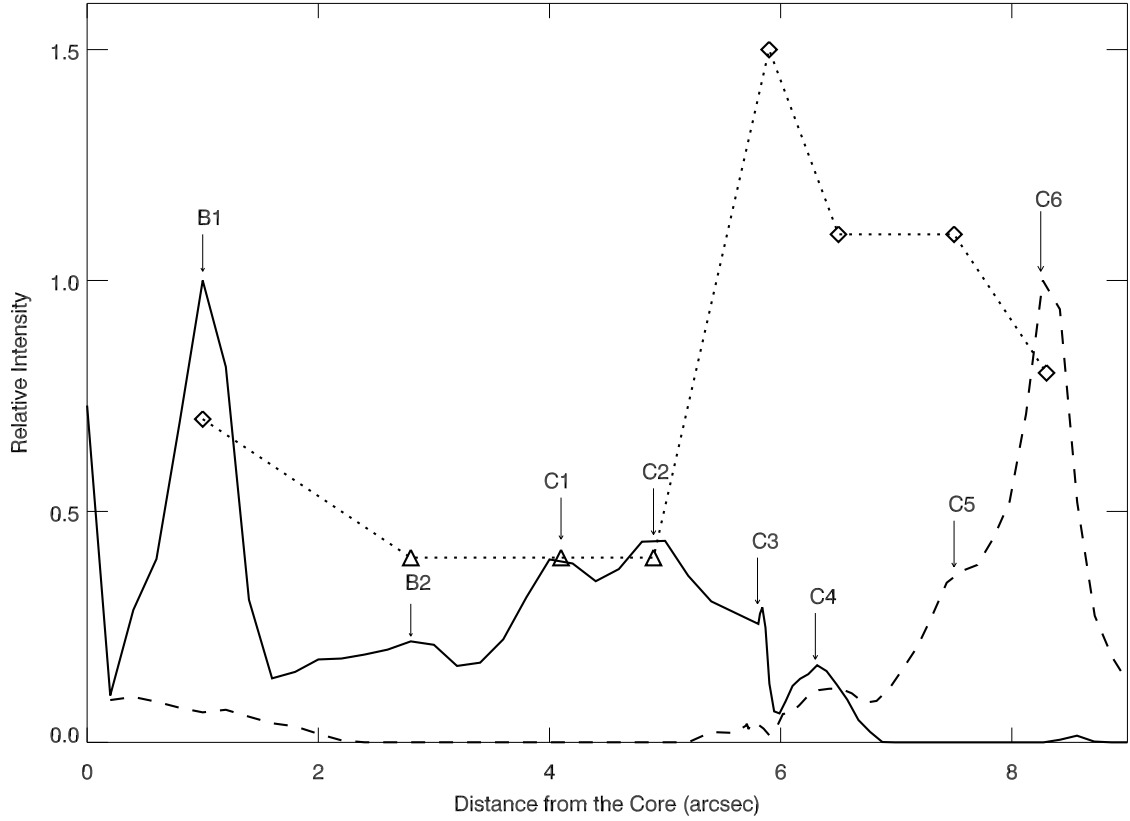}
\caption{The X-ray at 0.2-8~keV (solid line) and radio at 4.9~GHz (dashed line)
profiles of the jet along the axis indicated in Fig. 1. The dotted line
traces the spectral index along the jet in X-ray (triangles) and between
5 and 15~GHz (diamonds). \label{fig7}}
\end{figure}
\begin{figure}
\epsscale{1.0}
\plotone{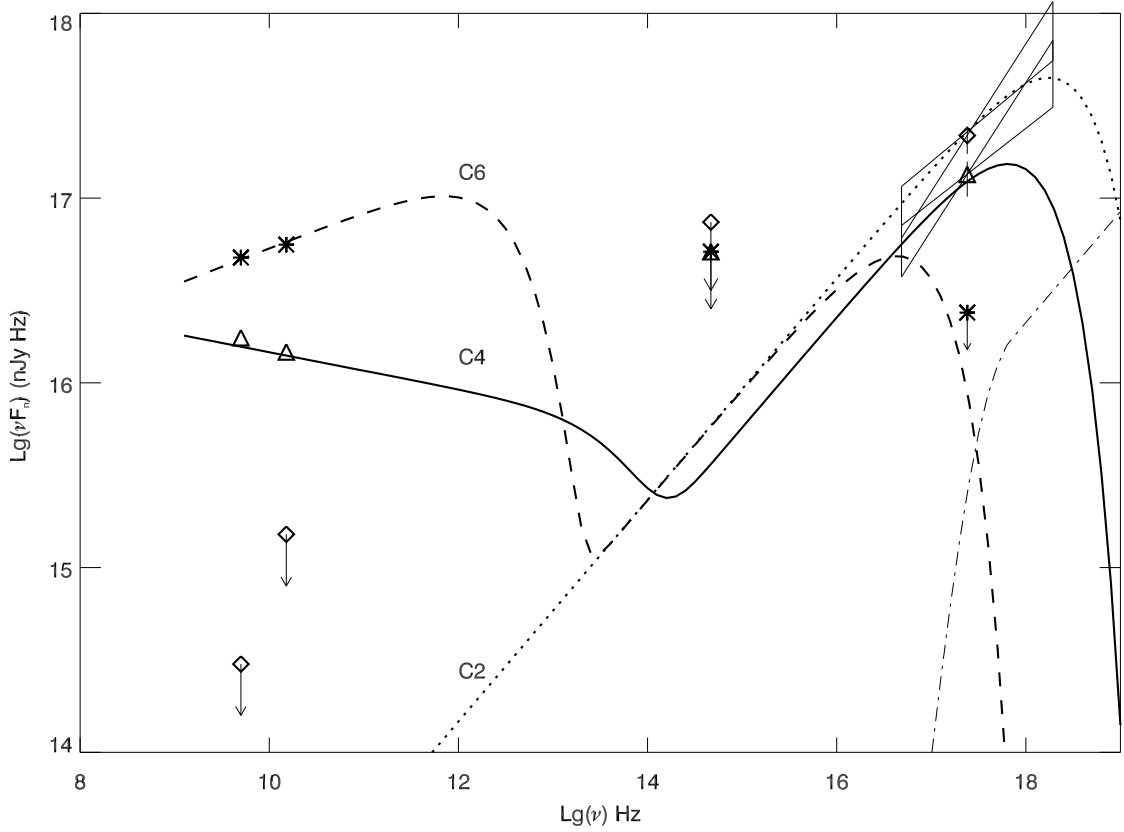}
\caption{Observed spectral energy distributions (SEDs) for knots $C2$ (diamonds),
$C4$ (triangles), and $C6$ (asterisks). Curves represent simulated SEDs for combined
synchrotron emission from a dual population of relativistic electrons:
dotted curve for $C2$, solid curve for
$C4$, and dashed curve for $C6$. The plasma parameters used to compute the model SEDs are listed
in Table 3. The dash-dot curve represents the simulated SED for IC/CMB emission
from the spine population with $\delta=20$ and $B=0.5$ $\mu$G.
\label{fig8}}
\end{figure}
\begin{figure}
\epsscale{1.0}
\plotone{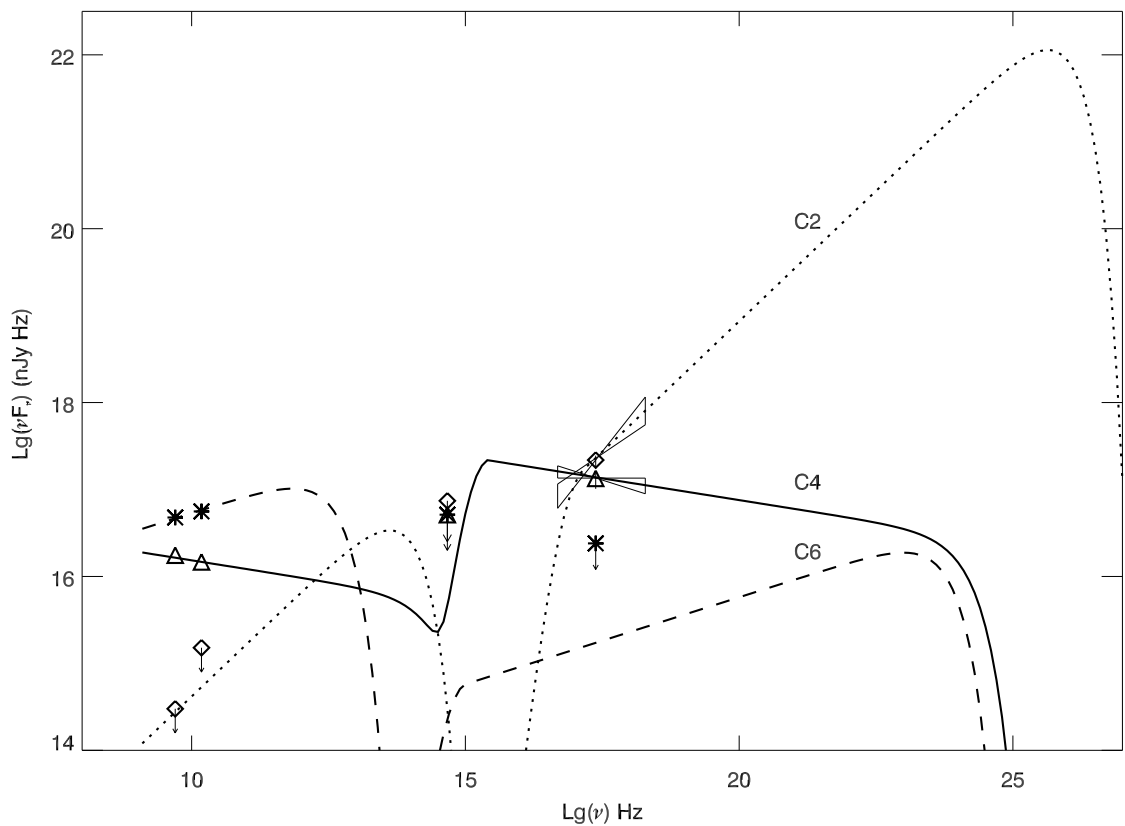}
\caption{The observed spectral energy distribution for knots $C2$ (diamonds),
$C4$ (triangles), and $C6$ (asterisks). The simulated SEDs for combined
synchrotron and CMB/IC emission of a single population of relativistic electrons with
$\gamma_{\rm min}=15$ and $\gamma_{\rm max}=5\times 10^5$ are shown by dotted curve for $C2$,
solid curve for $C4$, and dashed curve for $C6$. The
parameters used to compute the model SEDs are given in Table 4.
\label{fig9}}
\end{figure}
\begin{figure}
\epsscale{1.0}
\plotone{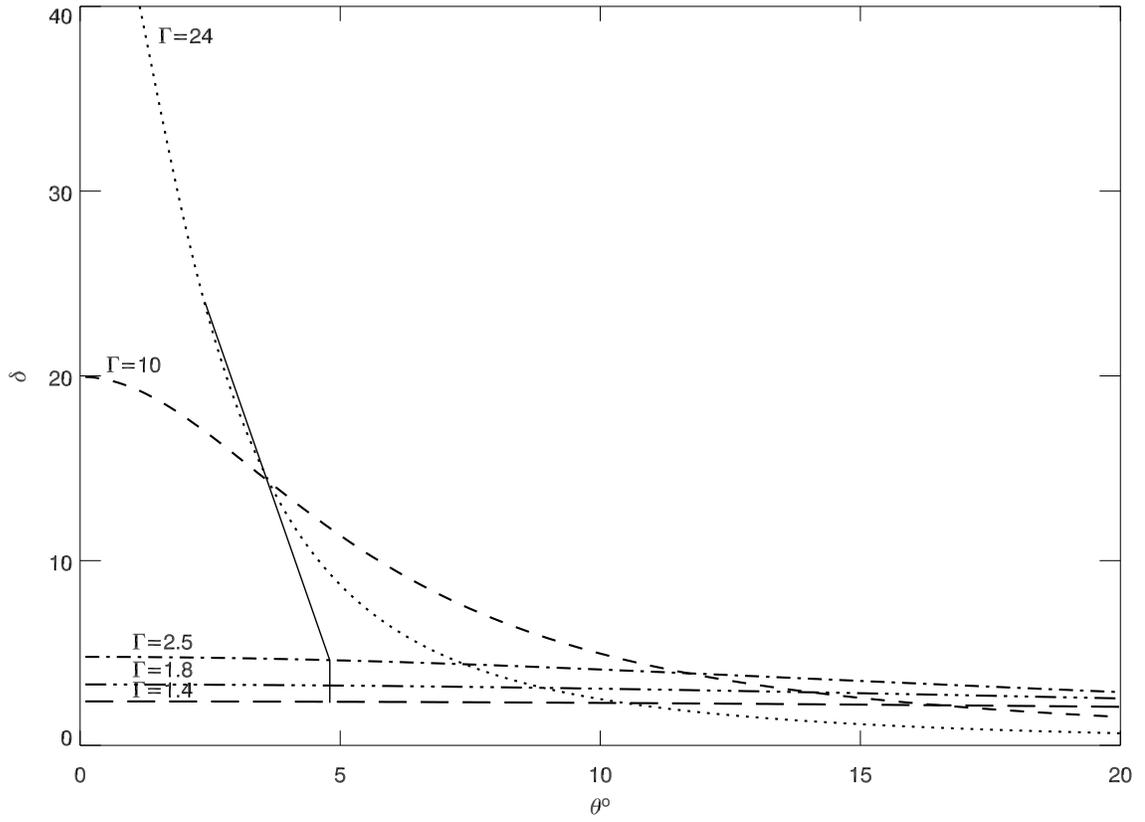}
\caption{Dependence of Doppler factor on viewing angle for different 
bulk Lorentz factors. Broken solid line indicates how the Doppler
factor might change from the core (24) to the outermost detected knot, $C6$ (2.3)
(See Table 4). 
\label{fig10}}
\end{figure}
\end{document}